\documentclass[twocolumn,preprintnumbers]{revtex4}

\input epsf

\usepackage[dvips]{color}

\usepackage{times}
\usepackage{graphicx}% Include figure files
\usepackage{dcolumn}% Align table columns on decimal point
\usepackage{bm}% bold math

\begin{document}

\title{\sf \Huge Evolution of Protein Interaction Networks \\ by
  Whole Genome Duplication \\ and Domain Shuffling}

\author{K.~Evlampiev \& H.~Isambert$^*$}

\vskip 0.2cm

\affiliation{ \vskip 0.2cm
\it Physico-chimie Curie, CNRS UMR168, Institut Curie, Section de
Recherche, 11 rue P. \& M. Curie, 75005 Paris, France.}%

\maketitle

\noindent
{\bf \small
Successive whole genome duplications have recently been 
firmly established in all major eukaryote kingdoms.
It is not clear, however, how such dramatic evolutionary
process has contributed to shape the large scale topology of 
protein-protein interaction (PPI) networks.
We propose and analytically solve a generic model of PPI network 
evolution under successive whole genome duplications.
This demonstrates that the observed scale-free degree distributions 
{\em and} conserved multi-protein complexes may have {concomitantly} arised
from  {\em i)} intrinsic {\em exponential} dynamics of
PPI network evolution and {\em ii)} {\em asymmetric divergence} of 
gene duplicates.
This requirement of asymmetric divergence is in fact ``spontaneously''
fulfilled at the level of protein-binding domains. 
In addition, domain shuffling of multi-domain proteins
is shown to provide a powerful  combinatorial source of PPI network
innovation, while preserving essential structures of the underlying
single-domain interaction network. Finally, large scale features of PPI
networks reflecting the ``combinatorial logic'' behind {\em direct} and 
{\em indirect} protein interactions are well reproduced numerically with only 
two adjusted parameters of clear biological significance. 
}

\small

\vspace{.4cm}

%%%%%%%%%%%%%%%%%%%%%%%%%%%%%%%%%%%%%%%%%%%%%%%%%%%%%%%%%%%%%%%%%%%%%%%%%%
\begin{figure*}
\includegraphics{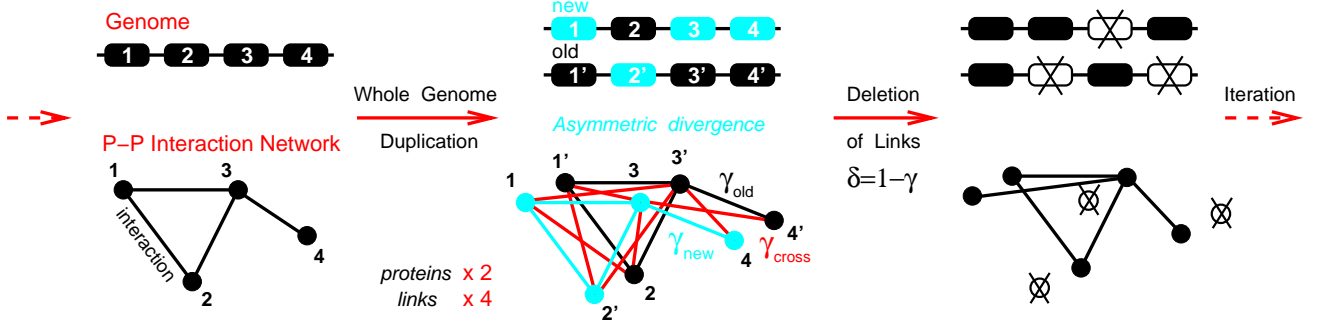}
\caption{\label{fig:wt} 
{\footnotesize
{\bf Model of protein-protein interaction network evolution 
through whole genome duplication.} 
Whole genome duplications are followed by {\em asymmetric} divergence
of protein duplicates with random distribution between genome copies 
({\it e.g.} 1/$1^\prime$ vs 2/$2^\prime$): ``New'' duplicates are  
left essentially free to accumulate neutral mutations
with the likely outcome to become nonfunctional and eventually deleted unless 
some ``new'', {\em duplication-derived} interactions are selected;
``Old'' duplicates, on the other hand, are more constrained 
to conserve ``old'' interactions already present before duplication.
The duplicated network with quadruplated 
links is graphically rearranged for convenience into old and new 
network copies ({\it e.g.} 2 and 2$^\prime$ duplicated nodes are swapped 
here). 
Links from the duplicated network are then kept with different 
probabilities $\gamma_i$ ($0\le\gamma_i\le 1$) reflecting this
asymmetric divergence between protein duplicates. 
An alternative model based on symmetric
  divergence of protein duplicates and random link 
``complementation''\cite{vazquez,middendorf}
is illustrated in Fig.S1 and discussed in the text.
}}
\vspace{-.5cm}
\end{figure*}
%%%%%%%%%%%%%%%%%%%%%%%%%%%%%%%%%%%%%%%%%%%%%%%%%%%%%%%%%%%%%%%%%%%%%%%%%%

%\vspace{0.3cm}
%\noindent
%{\bf \em Introduction}

Gene duplication is considered the main evolutionary source 
of new protein functions\cite{li}. Although long
suspected\cite{ohno,sparrow}, whole genome duplications have only been 
recently confirmed\cite{simillion,kellis,dujon,jaillon,dehal} 
through large scale comparisons of complete genomes\cite{wong,kellis}.

Whole genome duplications are rare evolutionary transitions  followed by
{random} nonfunctionalization of most gene duplicates on time scales
of about 100MY (with large variations between genes, see discussion).
Whole genome duplications presumably provide unique opportunities to 
evolve many new functional genes at once through accretion of functional 
domains\cite{doolittle1,riley,koonin,apic,orengo}  
from contiguous pseudogenes (or redundant genes)
and may also promote speciation events by preventing
genetic recombinations between 
close descendants with different random deletion patterns.

Recent whole genome duplications  (WGDs) within the last 500MY 
(about 15\% of life history) have now been firmly established 
in all major eukaryote kingdoms. 
For instance, there are 4 {consecutive} WGDs between 
the seasquirt {\it Ciona intestinalis} and the common carp
{\it Cyprinus carpio}, 
with most tetrapods (including mammals) in between 
at $+2$WGDs from seasquirt and $-2$WGDs from carp and most bony fish at
$+3$WGDs from seasquirt and $-1$WGDs from carp (a pseudotetraploid bony 
fish duplicated about 10MY ago)\cite{panopoulou,dehal,jaillon,david}.
There are also 3 {consecutive} WGDs in the recent evolution of the
flowering plant {\it Arabidopsis thaliana}\cite{simillion} and at least 3 
{consecutive} WGDs for the protist {\it Paramecium tetraurelia} (Patrick
Wincker,  personal communication). Extrapolating these 500MY old records, 
one roughly expects a few tens {\em consecutive} WGDs (or equivalent
``doubling events'') since the origin of life.
These rare but dramatic evolutionary transitions must have had
major consequences on the evolution of large biological networks,
such as protein-protein interaction (PPI) networks.

From a theoretical point of view, we also  expect that alternating
whole genome duplications and extensive gene deletions
lead to {\it exponential} dynamics of PPI network evolution.
In the long time limit, this should outweigh {all} {\it time-linear} 
dynamics that have been assumed in  PPI network evolution models
under {local} structure
changes\cite{albert2001,barabasi,raval,vazquez,berg,ispolatov1,ispolatov2}  
(see discussion). 
In fact, the intrinsic {exponential} dynamics of genome evolution is already 
transparent from the wide distribution of genome sizes\cite{li,sparrow}
and proliferation of repetitive elements\cite{hartl}:
it is  hard to imagine that the \mbox{$10^4$-fold} span
in lengths of eukaryote genomes could have solely
arised through time-linear increases (and decreases) in genome
sizes.*
%\footnote[3]{There is even a $10^5$-fold span in genome lengths when 
%including prokaryotes and $10^8$-fold including viruses!}.

\vspace{0.2cm}
\noindent
{\bf \em Modelling PPI network evolution by whole genome duplication}

We propose a simple model of PPI network evolution focussing on 
whole genome duplication (extensions to local or partial genome duplication
are presented in ref\cite{evlampiev_qbio} and confirm the conclusions of this
paper). Each time step $n$ corresponds to a whole genome
duplication and leads to a complete duplication of the PPI 
network, whereby each node is duplicated ($\times 2$)
and each interaction quadruplated  ($\times 4$) as depicted on Fig.1.
Links from the duplicated network are then kept with different 
probabilities $\gamma_i$ \mbox{($0\le\gamma_i\le 1$)} reflecting 
symmetric or asymmetric divergences between protein or link copies.

The interaction network is caracterized at each step $n$ by 
its number of nodes with $k$ neighbours $N_k^{(n)}$ and its total number of
links $L^{(n)}=\sum_{k\ge 1} k N_k^{(n)}/2$.
As stochastic differences exist between network realizations, we study the 
evolution of typical networks by  introducing a 
generating function averaged over all network realizations, 
\begin{equation}
\label{F_def}
F^{(n)}(x)=\sum_{k\ge 0} \langle N_{k}^{(n)}\rangle x^k.
\end{equation}
This use of generating functions can in fact be 
generalized\cite{evlampiev_qbio} to other, possibly non local features 
of interest ({\it e.g.} the average connectivity of first neighbors 
$g_k$\cite{maslov} is introduced below).  

In the following, we discuss a general model of PPI network evolution
through whole genome duplication with {\it asymmetric} divergence of
duplicated genes (Figs.1\&2A). We compare it, first, to an alternative 
model with {\it symmetric} protein divergence but random link 
``complementation''\cite{vazquez,middendorf} (Fig.S1),
and also to {\it direct} physical interactions from Yeast PPI network data
(Fig.~2B\&C). We then redefine this initial asymmetric divergence model 
(Fig.~1) in terms of protein-binding domains (Figs.~3A\&B) to account for
{\it indirect} protein-protein interaction within multi-protein complexes 
(Figs.~3A\&C).

\vspace{0.2cm}
\noindent
{\bf \em Asymmetric divergence of duplicated proteins}

The case of asymmetric divergence between duplicated genes corresponds to 
the following evolution scenario; while  duplicated proteins are initially  
equivalent and experience, at first, the same functional 
constraints\cite{kondrashov}, their divergence becomes eventually 
asymmetric\cite{zhang,conant,gu} (see discussion). 
This presumably occurs once one duplicate copy has lost an essential 
interaction and thus function, which has then to be fullfilled entirely
by the other duplicate. The evolution of this latter duplicate is, from
then on, more constrained to retain ``old'' interactions, while the former 
duplicate is left largely free to accumulate more neutral mutations
with the likely outcome to become nonfunctional, unless 
some ``new'', {\em duplication-derived} interactions are selected, Fig.~1
(new interactions arising from horizontal gene transfer are more 
characteristic of prokaryote evolution\cite{doolittle2} and neglected 
here\cite{ispolatov1}).
Note that ``old'' and ``new'' labels in Fig.~1  refer to the asymmetric 
conservation and fate of duplicates after WGD (and {\it not} to specific
genome copies).
Functionalization patterns of duplicated genes are further discussed in 
the supporting information.

The recurrence relation for the generating function (\ref{F_def}) is 
derived as follows: since each node is initially duplicated, $F^{(n+1)}(x)$ 
is the sum of two $F^{(n)}(x)$ where $x$ is first replaced by $x^2$ 
(since each node degree can at most double) and then substituted 
as $x\rightarrow \gamma_i x\!+\!\delta_i$ where $\gamma_i$
[resp. $\delta_i=1-\gamma_i$] corresponds to the probability to keep
[resp. delete]  each link emerging from each node of the duplicated graph.
Hence, the generating function recurrence for PPI network evolution with
asymmetric divergence of duplicated proteins yields, 
\begin{equation}
\label{F3}
F^{(n+1)}(x)=F^{(n)}\bigl((\gamma x\!+\!\delta)(\gamma_{\rm n} x\!+\!\delta_{\rm n})\bigr)+F^{(n)}\bigl((\gamma x\!+\!\delta)(\gamma_{\rm o} x\!+\!\delta_{\rm o})\bigr).
\end{equation}
where $\gamma$, $\gamma_{\rm n}$ and $\gamma_{\rm o}$ [resp. $\delta$,
$\delta_{\rm n}$ and $\delta_{\rm o}$] stand for $\gamma_{\rm cross}$,
$\gamma_{\rm new}$ and $\gamma_{\rm old}$ [resp.  $\delta_{\rm cross}$,
$\delta_{\rm new}$ and $\delta_{\rm old}$] in Fig.1  (see supporting
information for proof details). 

The overall graph dynamics through successive global duplications is clearly
exponential as anticipated; in particular, the total number of nodes grows as 
$F^{(n)}(1)=A\cdot 2^n$, where $A$ is the initial number of nodes, and
the number of links scales as $\langle
L^{(n)}\rangle \propto (2\gamma\!+\!\gamma_{\rm o}\!+\!\gamma_{\rm n})^n$.
We remove permanently disconnected nodes  from the list of relevant nodes,
assuming that they correspond to proteins that have in fact lost their
function and are eventually eliminated from the genome.  
To this end, we redefine the graph size as,
$\langle N^{(n)}\rangle=\sum_{k\ge 1} \langle N_k^{(n)}\rangle$
and introduce a normalized generating function $p^{(n)}(x)$ for 
the mean degree distribution, 
\begin{equation}
\label{p_def}
p^{(n)}(x)=\sum_{k\ge 1} p^{(n)}_k x^k,   {\rm ~~~~where~~~~} p^{(n)}_k={\langle N_{k}^{(n)} \rangle \over\langle N^{(n)}\rangle}.
\end{equation}
Absolute and relative generating functions are related through, 
\begin{equation}
\label{F-p}
F^{(n)}(x)=p^{(n)}(x)\langle N^{(n)}\rangle+\langle N_0^{(n)}\rangle.
\end{equation}
Inserting this expression (\ref{F-p}) in recurrence (\ref{F3}) gives  
a closed relation between successive $p^{(n)}(x)$, 
%%%%%%%%%%%%%%%%%%%%%%%%%%%%%%%%%%%%%%%%%%%%%%%%%%%%%%%%%%%%%%%%%
\begin{eqnarray}
\label{p3}
& &p^{(n+1)}(x)=1- \\ 
& &{2-p^{(n)}\bigl((\gamma x\!+\!\delta)(\gamma_{\rm n} x\!+\!\delta_{\rm n})\bigr)-p^{(n)}\bigl((\gamma x\!+\!\delta)(\gamma_{\rm o} x\!+\!\delta_{\rm o})\bigr)\over \Delta^{(n)}} \nonumber,
\end{eqnarray} 
where $\Delta^{(n)}$ is the ratio between consecutive numbers of connected
nodes, $\Delta^{(n)}={\langle N^{(n+1)}\rangle / \langle
  N^{(n)}\rangle}=2-p^{(n)}(\delta\delta_{\rm n})-p^{(n)}(\delta\delta_{\rm
  o}) \le 2$.

The evolution of the mean degree is obtained by taking the first derivative of
(\ref{p3}) at $x=1$:
\begin{equation}
\label{p4}
\partial_x p^{(n+1)}(1)={\Gamma_{\rm n}\!+\!\Gamma_{\rm o}\over \Delta^{(n)}} \partial_x p^{(n)}(1),
\end{equation}
where $\Gamma_{\rm n}=\gamma\!+\!\gamma_{\rm n}$ and $\Gamma_{\rm
  o}=\gamma\!+\!\gamma_{\rm o}$ hereafter.

We will limit the discussion here to degree distributions approaching a
stationary regimes $p^{(n)}(x) \rightarrow p(x)$ with a {\it finite} mean 
degree $1\le p'(1) < \infty$. This seems to cover the most biologically 
relevant networks; for completeness, other cases are discussed
elsewhere\cite{evlampiev_qbio}. 
From (\ref{p4}) and the condition of finite mean degree, we readily obtain
that $\Delta^{(n)}\rightarrow \Gamma_{\rm n}\!+\!\Gamma_{\rm o}$, 
which implies that the network evolution is asymptotically 
equivalent in terms of connected nodes and links,\dag
%\footnote{
%This condition can be shown\cite{evlampiev_qbio} to ensure 
%that the evolution of the {\em ensemble average} of networks 
%(Eq.\ref{F_def}) indeed reflects the ``typical'' 
%evolution of PPI networks under global duplication.},
\begin{equation}
\label{growthrate}
{\langle N^{(n+1)}\rangle / \langle N^{(n)}\rangle}\rightarrow{\langle
  L^{(n+1)}\rangle / \langle L^{(n)}\rangle}=\Gamma_{\rm n}\!+\!\Gamma_{\rm o} \le 2,
\end{equation}

The stationary degree distribution is then solution of the functional equation,
\begin{equation}
\label{p3func}
p(x)=1-{2-p\bigl((\gamma x\!+\!\delta)(\gamma_{\rm n} x\!+\!\delta_{\rm
    n})\bigr)-p\bigl((\gamma x\!+\!\delta)(\gamma_{\rm o} x\!+\!\delta_{\rm
    o})\bigr)\over \Gamma_{\rm n}\!+\!\Gamma_{\rm o}}, 
\end{equation}
which can be differentiated $k$ times to express the $k$th derivative in terms
of lower derivatives,
\begin{equation}
\label{derivatives2}
\partial_x^{k}p(1)\left[1-{\Gamma_{\rm n}^k+\Gamma_{\rm o}^k\over \Gamma_{\rm n}+\Gamma_{\rm o}}\right]=\sum_{m=[k/2]}^k \alpha_m(\gamma_{\rm n}, \gamma_{\rm o}, \gamma) \partial_x^{m}p(1)
\end{equation}
where the coefficients $\alpha_m$ are all positive from the definition (\ref{p_def}).

%%%%%%%%%%%%%%%%%%%%%%%%%%%%%%%%%%%%%%%%%%%%%%%%%%%%%%%%%%%%%%%%%%%%%%%%%%
\begin{figure*}
\includegraphics{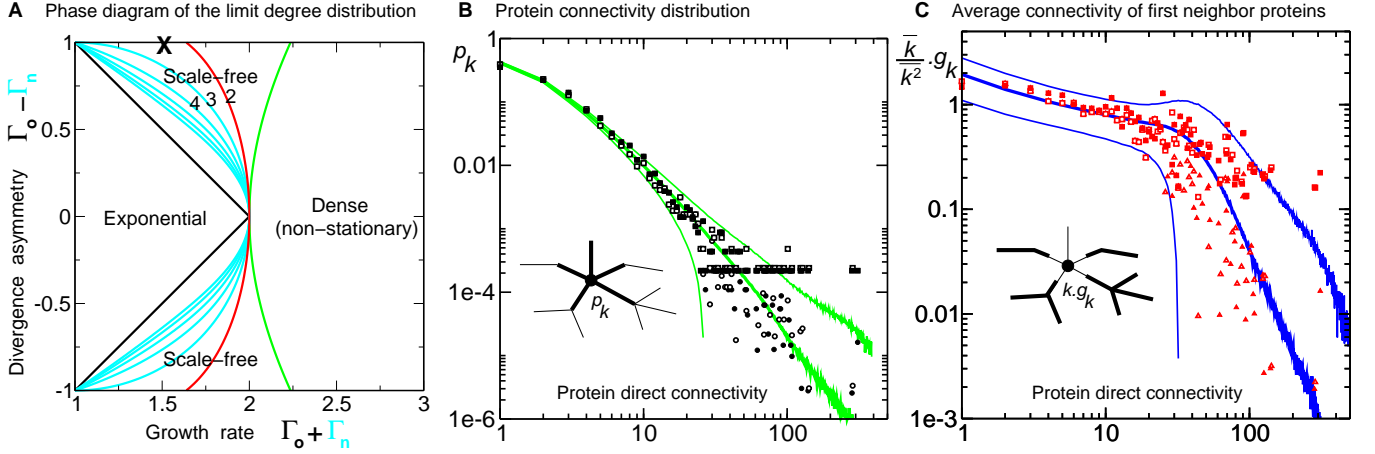}
\caption{\label{fig:wt} 
{\footnotesize 
{\bf Analytical and numerical results of PPI Network evolution through whole
  genome duplication.} 
{\bf A.} {Phase diagram for the limit degree distribution (see text).}
{\bf B\&C.} Comparison with protein {\it direct} physical interaction data for
Yeast from BIND\cite{bind} and MIPS\cite{mips} databases: BIND (August 11,
2005 release), 4576 proteins, 9133 physical interactions,  
$\overline{k} =3.99$, $\overline{k^2}=106$ (filled symbols) 
and MIPS (downloaded online April 20, 2006), 4153 proteins, 7417 physical
    interactions, $\overline{k} =3.57$, $\overline{k^2}=78.6$ 
(open symbols). Squares correspond to raw data, while circles and triangles
are statistically averaged with gaps in connectivity distribution for large 
$k\ge 20$, due to the finite size of Yeast PPI network. 
{\bf B.} {One-parameter fit of connectivity distribution data $p_k$
  (corresponding to the  ``{\sf X}'' mark in  {\bf A.}, see text).}
Numerical connectivity distribution averaged over 10,000 network realizations
(central green line). Numerical averages plus or minus two standard deviations
($\pm 2 \sigma$) are also displayed to show the predicted dispersions (upper
and lower green lines) [Raw data (squares) do not fit within the mean 
$\pm 2\sigma$ curves for large $k$ due to the finite size of Yeast PPI
network]. The fitting parameter $\gamma=0.26$ corresponds to an effective
growth rate of $1+2\gamma=1.52$. 
{\bf C.}  One-parameter fit of average connectivity of first neighbor proteins
$g_k$\cite{maslov} ({\it i.e.} $k.g_k$ sums connectivities of first neighbors
from proteins of connectivity $k$). Numerical predictions averaged over 10,000
network realizations (central blue line). Numerical averages plus or minus two
standard deviations are also displayed (upper and lower blue lines).  
Same fitting parameter value as in  {\bf B}, $\gamma=0.26$. 
Note that $g_k$ is rescaled by $\overline{k} / \overline{k^2}$ 
(as $\overline{k g_k}=\overline{k^2}$ holds for each network realization);
this rescales large $g_k$ fluctuations between network realizations, due to 
the divergence of $\overline{k^2}$ for $p_k \sim k^{-\alpha-1}$
with $2>\alpha>0$  for the one-parameter model.}} 
\vspace{-.5cm}
\end{figure*}
%%%%%%%%%%%%%%%%%%%%%%%%%%%%%%%%%%%%%%%%%%%%%%%%%%%%%%%%%%%%%%%%%%%%%%%%%%

The finite or infinite nature of $\partial_x^{k}p(1)$ depends on the two  
parameters $\Gamma_{\rm n}$ and $\Gamma_{\rm o}$ and defines the form of the
limit degree distribution. The phase diagram Fig.~2A summarizes in the plane
$(\Gamma_{\rm o}\!+\!\Gamma_{\rm n}, 
\Gamma_{\rm o}\!-\!\Gamma_{\rm n})$ the different regimes for the asymptotic
degree distribution $p_k$. $\Gamma_{\rm o}\!+\!\Gamma_{\rm n}$ is
the {\em global growth rate} of the network ($\Gamma_{\rm o}\!+\!\Gamma_{\rm
  n}>1$ to ensure a growing network) and $\Gamma_{\rm o}\!-\!\Gamma_{\rm n}$
corresponds to the {\em divergence asymmetry} between  duplicated proteins.
We now discuss the two main stationary regimes for $p_k$ in the case of 
$\Gamma_{\rm n}\le\Gamma_{\rm o}$ (the case $\Gamma_{\rm n}\ge\Gamma_{\rm o}$
is deduced by permutating indices):\\

\noindent
$\bullet$ {\em Exponential, non-conservative regime.} If both $\Gamma_{\rm
  o}\!<\!1$ and $\Gamma_{\rm n}\!<\!1$, %then,
\begin{equation}
\Gamma_{\rm n}^k+\Gamma_{\rm o}^k<\Gamma_{\rm n}+\Gamma_{\rm o}, {\rm  ~~~~for~ all~~} k\geq 2
\end{equation}
and the factor in
front of $\partial_x^{k}p(1)$ in (\ref{derivatives2}) is always strictly
positive, which implies that all derivatives of the limit degree
distribution are finite. Hence, in this case, the limit degree distribution 
decreases more rapidly than any power law (see explicit asymptotic 
development in\cite{evlampiev_qbio}). 
Note that this ``exponential'' regime occurs when the {\it links emerging 
from each node (Fig.~1) are more likely lost than duplicated} at each round 
of global  duplication  
(as $\Gamma_{\rm i}=\gamma\!+\!\gamma_{\rm i}<1$ is equivalent to
$\delta\delta_{\rm i}>\gamma\gamma_{\rm i}$). This implies that most nodes 
eventually disappear, and with them all traces of network
evolution,  after just a few rounds of global duplication. 
The network topology is {\it not} conserved, but instead continuously renewed
from duplication of the (few) most connected nodes.\\ 

\noindent
$\bullet$ {\em Scale-free, conservative regime.} If $\Gamma_{\rm o}>1>\Gamma_{\rm n}$, %then
the factor in front of $\partial_x^{k}p(1)$ in (\ref{derivatives2}) can become
negative. However, since the 
generating function should have all its derivatives positive, a negative value
for one of them means that it simply does not exist. In fact, for $\Gamma_{\rm
  n}\ln\Gamma_{\rm n}\!+\!\Gamma_{\rm o}\ln\Gamma_{\rm o}\ge0$ (red line in
Fig.2A and \cite{evlampiev_qbio}), there is an integer $r\ge1$ such that,
\begin{equation}
\Gamma_{\rm n}^{r}+\Gamma_{\rm o}^{r}\leq \Gamma_{\rm n}+\Gamma_{\rm o}<\Gamma_{\rm n}^{r+1}+\Gamma_{\rm o}^{r+1}.
\end{equation}
implying that all derivatives $\partial_x^{k}p(1)$ are finite up to the $r$th
order, while $\partial_x^{r+1}p(1)$ is infinite. This justifies the following
asymptotic expansion of $p(x)$ in the vicinity of $x=1$,
\begin{equation} 
\label{anzats}
p(x)=1-A_1(1-x)+\ldots+(-1)^r A_r(1-x)^r-A_{\alpha}(1-x)^{\alpha}-\ldots,
\end{equation}
for some appropriate $r<\alpha<r+1$. This anzats is then inserted in
(\ref{p3func}) using $(\gamma x\!+\!\delta)(\gamma_{\rm n,o} x\!+\!\delta_{\rm
  n,o})=1-\Gamma_{\rm n,o}(1-x)\!+\!\gamma\gamma_{\rm n,o}(1-x)^2$ 
to obtain an equation on the coefficients $A_1$,... $A_r$. The term $A_\alpha$ 
does not mix with previous terms and gives the following equation for $\alpha$,
\begin{equation}
\label{eq_alpha}
\Gamma_{\rm n}^{\alpha}+\Gamma_{\rm o}^{\alpha}=\Gamma_{\rm n}+\Gamma_{\rm o}.
\end{equation}
The limit degree distribution follows a power law in this case,\ddag
%\footnote{When 
%$\Gamma_{\rm n}^r+\Gamma_{\rm o}^r=\Gamma_{\rm n}+\Gamma_{\rm o}$
%for  exactly some integer $r\ge1$ the last term of the asymptotic formula
%  (\ref{anzats}) should be replaced by $(1-x)^r\ln(1-x)$, and the limit degree
%  distribution  decreases like $k^{-r-1}$.}, 
\begin{equation}
 p_k\propto k^{-\alpha-1},
\end{equation}
(see red and blue ``exponent'' lines  in Fig.~2A for $\alpha\!+\!1=2, 3,
4,\ldots$) 

Note that scale-free degree distributions emerge under successive, global
network duplications only if {\it the ``old'' node copy has its links
more likely duplicated than lost} at each round of global duplication (as
$\Gamma_{\rm o}=\gamma\!+\!\gamma_{\rm o}>1$ is equivalent to
$\gamma\gamma_{\rm o}>\delta\delta_{\rm o}$).
Thus, ``old'' nodes statistically keep on increasing their connectivity
once they have emerged as ``new'' nodes by duplication.
This implies that most nodes {and} their surrounding links are conserved 
{\it throughout} the evolution process, thereby ensuring that local 
topologies of previous networks remain embedded in subsequent networks. 

In summary, whole genome duplication with asymmetric divergence of 
duplicated proteins leads to the emergence of two classes of PPI networks 
with finite asymptotic degree distributions :
{\em i)} PPI networks with an exponential degree distribution and without
conserved topology  
and {\em ii)} PPI networks with a scale-free limit degree distribution and at
least local topology conservation.  
All other evolution scenarios, which do not lead to finite asymptotic degree 
distributions, are unlikely to model biologically relevant cases; 
they correspond  either to an {\em exponential} disappearance of the whole 
PPI network ({\it i.e.} if $\Gamma_{\rm n} + \Gamma_{\rm  o}<1$) or 
to an {\em exponential} shift of {\it all} proteins towards higher and higher
connectivities ({\it i.e.} dense regime in Fig.~2A 
for $\Gamma_{\rm n}\Gamma_{\rm  o}>1$)\cite{evlampiev_qbio}.  

\newpage

\vspace{0.2cm}
\noindent
{\bf \em Symmetric divergence of duplicates with link ``complementation''}

Another model of interest is the 
so-called ``duplication-mutation-complementation''
 model initially proposed in the context of protein network evolution through
 successive {\it local} duplications\cite{vazquez,middendorf}. 
This model can be easily adapted to the 
context of PPI network evolution through whole genome duplication, Fig.~S1.
After each global duplication step, the probability to keep an instance
of each interaction is now distributed randomly over the four equivalent links
without reference to particular protein duplicates, unlike in the previous
model.  
The complementation step (which ensures that at least one instance of each
previous link is retained)
can be enforced here through the ``old'' link copy ($\gamma_{\rm o}=1$) with 
$\gamma_{\rm n}$ corresponding to the ``new'' interaction sharing no node with 
 $\gamma_{\rm o}$, while $\gamma$ still pertains to the last two equivalent
 cross  links. 
 This model is thus effectively symmetric from the protein point of view and
 readily yields the following recurrence for the generating function of the
network degree distribution.
\begin{equation}
\label{F3sym}
F^{(n+1)}(x)=2 F^{(n)}\bigl((\gamma x\!+\!\delta)({\gamma_{\rm
    e}} x\!+\!{\delta_{\rm e}})\bigr),
\end{equation}
where $\gamma_{\rm e}=(\gamma_{\rm n}\!+\!\gamma_{\rm o})/2$ and $\delta_{\rm
  e}=(\delta_{\rm n}\!+\!\delta_{\rm o})/2$ are effective average probabilities
  to retain or delete old and new links (see supporting information for proof
  details). Hence, the model of PPI network evolution with link
  complementation is in fact equivalent to the case of a  symmetric divergence
  of duplicated proteins in the previous general model.
Such symmetric divergence of duplicated proteins yields either a stationary
exponential regime ($\Gamma_{\rm n}+\Gamma_{\rm o}<2$, Fig.2A) or a
non-stationary dense regime\cite{evlampiev_qbio} 
($\Gamma_{\rm n}+\Gamma_{\rm o}>2$, Fig.2A). 

Hence, the ``duplication-mutation-complementation'' model
{\it cannot} lead to scale-free degree distributions, and thus to locally
conserved network topology, in the context of whole
genome duplication evolution, by contrast to the same model
applied to local duplication with time-linear 
evolution\cite{vazquez,middendorf}.

\vspace{0.2cm}
\noindent
{\bf \em Fitting PPI network data with a one-parameter model}

Scale-free degree distributions have been widely
reported for large biological networks and other exponentially growing
networks like the WWW. We showed in the previous discussion that 
scale-free limit degree distributions require an asymmetric divergence
of duplicated proteins ($\Gamma_{\rm o}-\Gamma_{\rm n}=\gamma_{\rm
  o}-\gamma_{\rm n}>0$) which corresponds to the probability difference
between conservation of old interactions ($\gamma_{\rm o}$) 
and coevolution of new binding sites ($\gamma_{\rm n}$).
The expected range of parameters for actual biological networks 
is $1 \simeq \gamma_{\rm o} \gg \gamma\gg \gamma_{\rm n} \simeq  0$;
In particular,  the most conservative ($\gamma_{\rm o}=1$) and least
correlated ($\gamma_{\rm n}=0$) evolution scenario corresponds to the
strongest divergence asymmetry between duplicated proteins 
($\Gamma_{\rm o}-\Gamma_{\rm n}=1$, upper border on Fig.2A).
The condition $\gamma_{\rm o}=1$ ensures that not only local but also
global topologies  of all previous networks  remain embedded 
in all subsequent networks. This model is effectively a one-parameter model
($\gamma$) for PPI network evolution through whole genome  duplication. It
converges towards a stationary scale-free limit degree distribution $p_k \sim
  k^{-\alpha-1}$ with $0<\alpha<2$ for $0<\gamma<(\sqrt{5}-1)/2$ and generates
  non-stationary dense  networks for
  $(\sqrt{5}-1)/2<\gamma<1$\cite{evlampiev_qbio}.    
We used this one-parameter model to fit both the degree distribution (Fig.2B)
and the  average connectivity of first neighbors (Fig.2C) 
for {\it direct} physical interaction data of {\it S. cerevisiae}
taken from two databases, BIND\cite{bind} %(http://www.bind.ca) 
and hand curated %(http://www.mips.com)  
MIPS\cite{mips} (with presumabky fewer nonspecific spurious 
interactions\cite{deeds}). The predicted asymptotic regime is in fact
approached for $k\le 20$ due to the finite size of Yeast PPI network. 
The fitting parameter $\gamma=0.26$ corresponds to a fixed growth rate
(\ref{growthrate}) of $1+2\gamma=1.52$  ({\it i.e.} the number of links
and nodes increases by 52\% at each global duplication).
Adding and removing up to 30\% of links randomly, or
drawing $\gamma$ from a uniform distribution between 0 and 0.52 (with average
$\bar{\gamma}=0.26$) yield  remarkably similar fits (not shown) to the
experimental data. This reveals a large insensibility to false- positive and
negative noises and fluctuations in $\gamma$
(as long as the non-stationary dense regime is avoided, Fig.2A).
 The fixed (or averaged) growth rate of 52\% at each round of global
 duplication is enough to generate  
networks of the size of {\it S. cerevisiae} starting  from a few interacting
``seeds'' after about 20 global
duplications ({\it i.e.} $1.52^{20}=4334$ times more nodes with an average
of one global duplication per 200MY for 4BY). 
Such scenario is not {\it a priori} incompatible with experimental data,
as we only have clear records on global duplications dating back up to
400-500MY ago ({\it i.e.} only 10 to 20\% of life history).  
Yet, these records suggest that ``recent'' whole genome duplications
might be more frequent (every 100-150MY) and more selective (growth
rates between 10 and 25\%).\S
%\footnote{{\it Ciona} (16,000 genes) and {\it
%    human} [resp. {\it
%    tetraodon}]  ($\sim$22,000 genes)  differ by two [resp. three] whole
%    genome duplications; %  dating back more  than 300MY; 
%    this corresponds to an averaged growth rate
%    of 17\% [resp. 11\%].}. 
%{\it i.e.}  $(22/16)^{1/2}=1.17$ ~~[resp. $(22/16)^{1/3}=1.11$].}.

\vspace{0.2cm}
\noindent
{\bf \em Direct vs indirect protein-protein interactions}

The protein-protein interactions we have considered so far correspond to
{\it direct} physical contact between {\it protein pairs} derived, for
instance, from two-hybrid expression assays\cite{fields}.
However, we expect from the proposed scale-free fit of the
degree distribution (Fig.~2B) that the underlying PPI network has conserved
not only pairwise interactions during evolution but also some level of network
topology (see above). The emergence of locally conserved topology in PPI
network evolution leads ``naturally'' to conserved associations or ``modules''
between multiple
proteins\cite{dokholyan,spirin,wuchty2003,wuchty2004,vergassola} and, 
beyond, to recurrent ``motifs'' across different types of biological
networks\cite{hartwell,milo2002,guelzim,yeger-lotem2004,francois,berg2,mazurie,buchler2005}. 

In fact, many biological functions are known to rely on 
multiple direct and indirect interactions within protein complexes. Moreover,
the {\it combinatorial} complexity of multiple-protein interactions 
is likely responsible for the remarkable diversity amongst living 
organisms\cite{birchler}, despite their rather limited and largely 
shared genetic background
({\it i.e.} a few (ten) thousands genes built from a few hundreds 
families of homologous protein domains\cite{murzin,apic,superfamilly,orengo}).

High-throughput studies using affinity precipitation methods
coupled to mass spectroscopy\cite{gavin2002,ho2002,gavin2006} have  
proposed some 80,000 direct and indirect protein interactions for 
{\it S. cerevisiae} (raw data) and similar data are now becoming 
available for several other species. 

Yet, from a theoretical point of view, the evolution of {\it indirect} 
interactions is expected to depend not only on locally conserved network
topology but also on the actual ``combinatorial logic'' between direct
interactions. This cannot be readily defined on traditional PPI network
representation ({\em e.g.} Fig.~1) and requires a somewhat more elaborate 
model as we now discuss.  

%%%%%%%%%%%%%%%%%%%%%%%%%%%%%%%%%%%%%%%%%%%%%%%%%%%%%%%%%%%%%%%%%%%%%%%%%%
\begin{figure*}
\includegraphics{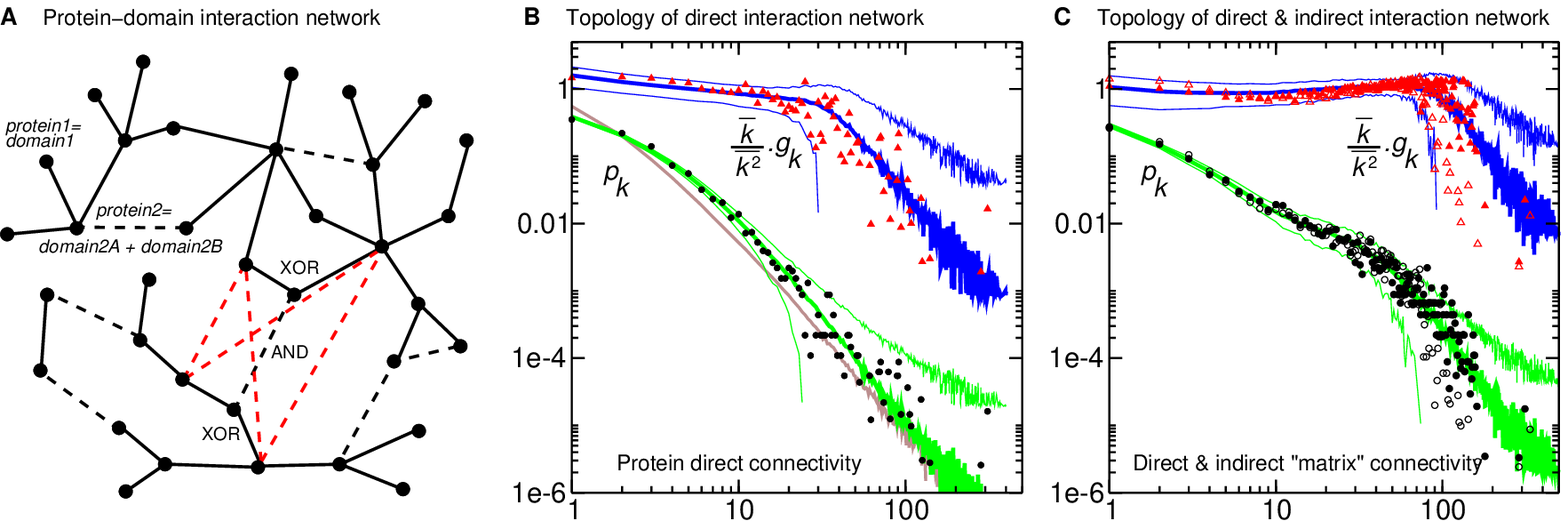}
\caption{\label{fig:wt} 
{\footnotesize
{\bf Combining whole genome duplication and domain shuffling of multi-domain
  proteins.}\\ 
{\bf A.} {Protein-domain interaction network.}
Nodes now correspond to single binding domains in a protein-domain 
interaction network (solid lines).  
Multi-binding-domain proteins are introduced through a new type of 
links corresponding to covalent peptide bonds between protein domains 
(black dashed lines). This provides a graphical framework to distinguish  
 mutually exclusive, direct interactions (``XOR'') between protein domains
 from  cummulative, indirect interactions (``AND'') within multi-protein
 complexes (red dashed lines).
{\bf B\&C.} Comparison with protein direct \& indirect interaction data 
for Yeast from BIND\cite{bind} 
database ({\bf B\&C} filled symbols, indirect interactions
from\cite{ho2002,gavin2002}) and Ref\cite{gavin2006} ({\bf C} open symbols,
see supporting information). Data are statistically averaged as in Fig.~2B\&C
to account for gaps in connectivities for large $k\ge 20$,  
due to the finite size of Yeast PPI network. {\bf B.} Two-parameter fit of
both direct connectivity distribution $p_k$ and average direct connectivity of
first neighbor proteins $g_k$\cite{maslov} (see Fig.2C and text). 
Numerical predictions are averaged over 1,000 network realizations 
(central green and blue lines). Numerical averages plus or minus two standard
deviations are also displayed to show the predicted dispersions (upper and
lower green and blue lines). The two adjusted parameters ($\gamma=0.1$ and
$\lambda=0.3$) correspond to a network growth rate of 20\% and an average of 
1.5 protein-binding sites (domains) per protein. The connectivity distribution
of the underlying single-domain network (corresponding to $\gamma=0.1$ and 
$\lambda=0.0$) is also displayed (brown line) to illustrate its 
relation to the full multi-domain protein network (see text).
{\bf C.} Two-parameter fit of both direct \& indirect ``matrix'' connectivity 
distribution $p_k$ and average direct \& indirect ``matrix'' connectivity of
first neighbor proteins $g_k$\cite{maslov} (see text). 
Same two adjusted parameters ($\gamma=0.1$ and
$\lambda=0.3$) as in {\bf B} while a selection of indirect interactions 
is added up to a total of 28,000 direct \& indirect
interactions (see supporting information).}}  
\vspace{-.5cm}
\end{figure*}
%%%%%%%%%%%%%%%%%%%%%%%%%%%%%%%%%%%%%%%%%%%%%%%%%%%%%%%%%%%%%%%%%%%%%%%%%%

\vspace{0.2cm}
\noindent
{\bf \em Redefining PPI network evolution in terms of protein domains}

Indirect protein interactions reflect the occurence 
of  {\it simultaneous} direct interactions within protein complexes.
This requires that some proteins have more than one binding
sites to simultenaously interact with several protein partners.
Indeed, proteins with a single protein-binding site can only bind to one 
partners at a time, underlying a simple ``XOR''-like  combinatorial logic.
By contrast, proteins with several protein-binding sites (which are usually
multi-domain proteins) greatly increase the combinatorial complexity of
biological processes (like gene regulation or cell signaling) by adding 
``AND'' operators to the computational logic between multiple direct
interactions. Multi-domain proteins also provide a versatile support for
protein evolution through accretion or deletion of individual
domains\cite{doolittle1,riley,koonin,apic,orengo}. 

In addition, we note that binding sites\cite{sheinerman,levy} on specific 
protein domains are likely the primary source of asymmetric divergence in 
PPI network evolution, as binding site mutations necessarily affect 
interactions with {\it all} binding partners (Fig.~1) and not just a random 
subset of them (Fig.~S1). Hence, asymmetric divergence of protein duplicates 
``naturally'' originates from ``spontaneous symmetry breaking'' of their
equivalent protein-binding sites (or domains).

We propose to highlight this central role of protein domains in the evolution
of PPI networks by simply redefining our initial asymmetric divergence model
(Fig.~1) in terms of {\it protein-binding domains} ({\it i.e.} with a single
protein-binding site) as illustrated in Fig.~3A. This alternative
representation of PPI networks provides a theoretical framework to model the
evolution of the combinatorial logic  underlying PPI networks, as it
distinguishes  mutually exclusive, direct interactions (``XOR'') between
protein domains  (Fig.3A, black solid lines) from cummulative, indirect
interactions (``AND'') within multi-protein complexes (Fig.3A, red dashed
lines). 

\vspace{0.2cm}
\noindent
{\bf \em Combining whole genome duplication and domain shuffling.}

As noted in the introduction, whole-genome duplications promote efficient 
shuffling of multi-domain proteins by enabling many accretion and deletion
events of functional domains after each genome doubling. We will assume in the 
following that this shuffling of multi-domain proteins is so efficient that
protein {\it domains} encoded along the genome evolve {\it
  independently}  from their inclusion in single- or multi-domain
proteins (indeed, different multi-domain combinations are typically 
observed across living kingdoms\cite{orengo}).
Besides, a more  elaborate model of protein evolution detailing domain 
accretion and deletion events leads to virtually identical results for the 
large scale topological features of PPI network (not shown). 
The asymptotic generating function $\tilde p(x)$ for multi-domain 
protein networks with {\it independent} domain evolution 
can be deduced {\it a posteriori} as,
\begin{eqnarray}
\label{pMD}
\tilde p(x)\!=\!(1-\lambda)p(x) \bigl( 1+ \lambda p(x)  + \lambda^2 p^2(x) +
\ldots \bigr)\!=\!{(1-\lambda)p(x) \over 1 - \lambda p(x) }\nonumber 
\end{eqnarray} 
where $\lambda$ is the probability of covalent connection
between successive protein domains encoded along the genome. 
This leads to an exponential distribution of multi-domain proteins, in 
agreement with actual distributions\cite{wolf,ekman}, with an average 
of $1/(1-\lambda)$ protein-binding sites per protein.  
While $p(x)$ now reflects the independent evolution
of single protein-binding domains according to Eqs.(\ref{p3func},\ref{anzats}),
it also controls the asymptotic properties of the derived multi-domain
networks $\tilde p(x)$; in particular, for $\Gamma_{\rm
  o}>1>\Gamma_{\rm n}$, we obtain from Eq.(\ref{anzats}) the following
asymptotic expansion in the vicinity of $x=1$, 
\begin{eqnarray}
\label{pMD}
& &\tilde p(x)=1-{1-p(x) \over 1 -
  \lambda p(x)} \sim  1 -\ldots - {A_{\alpha} \over
  1-\lambda}(1-x)^{\alpha}-\ldots \nonumber 
\end{eqnarray} 
which implies that degree distributions of multi-domain protein networks
$\tilde p_k$ increase with respect to the underlying single-domain
interaction network $p_k$ as $\tilde p_k \sim p_k  /(1-\lambda)$ for 
large $k$, while the fraction of proteins with a single binding partner 
$\tilde p_{1}$ decreases at the same time as  
$\tilde p_1= \tilde p^\prime(0) = (1-\lambda) p^\prime(0)=(1-\lambda)p_1$ 
(see Fig.~3B). Note that the scale-free degree distribution of such
multi-domain protein networks results from an {\it asymmetric 
divergence of individual binding sites} (or domains) rather than asymmetric 
divergence of global  protein architectures. This has also consequences for
the functionalization of duplicated genes (see supporting information).
In particular,  random (symmetric) ``subfunctionalization'' between protein  
duplicates {\it at the level of protein domains} does {\it not} prevent the 
emergence of scale-free networks with locally conserved topology, by
contrast to  random link ``complementation''  {\it at the level of 
individual interactions} (Fig.~S1) which leads to exponential
networks without  conserved topology (as discussed above).   
 
Hence, domain shuffling of multi-domain proteins provides a powerful, 
yet non-disruptive source of combinatorial innovation, as it preserves 
essential topological features inherited from the underlying protein-domain 
interaction network evolution. 

Finally, comparison with experimental data sets including indirect 
protein-protein interactions\cite{gavin2002,ho2002,gavin2006} is made
by adopting a statistical implementation of the ``combinatorial logic''
discussed above (see supporting information). It is based on a Dijkstra 
algorithm that estimates the relative importance of all possible indirect
interactions between multi-domain (and single-domain) proteins 
for each PPI network realization. Figs.~3B\&C show  rather good fits 
of experimental data sets corresponding to an estimated 30\% to 60\%
coverage of actual PPI networks\cite{gavin2002,ho2002,gavin2006}
(see, however, supporting information). The two adjusted parameters,
$\gamma=0.1$ and $\lambda=0.3$, correspond to a network growth rate of 20\% 
({\it i.e.} $1+2\gamma$) and an average of 1.5 ({\it i.e.} $1/(1-\lambda)$) 
protein-binding sites (domains) per protein in agreement with broad estimates 
for these biological parameters (see above \S~ and \cite{wolf,ekman}).
This also confirms that the properties of PPI networks we have predicted
from first principles ({\it i.e. i)} exponential dynamics and {\it ii)}
symmetry breaking) are already transparent from partial data sets.

\vspace{0.2cm}
\noindent
{\bf \em Discussion}

Beyond whole genome duplications, {\it local} genome rearrangements such as
 small segmental duplications, rearrangements and horizontal transfers  might
 well have been critical for the emergence and proliferation of living
 organisms. Moreover, we note that local duplications/deletions may also lead
 to exponential dynamics of PPI network evolution if they are selected
 independently in parallel (exponential models of local or partial genome
 duplication are presented in ref\cite{evlampiev_qbio}). Yet, recent records
 ($<$500MY) from various eukaryote kingdoms (from protists to animals and
 plants) suggest that the majority of duplicates may still have arised from  
successive whole genome duplications (although this will need to be confirmed
as more fully sequenced eukaryote genomes will become available).

One possible origin for this less efficient selection of local duplications
might be the dosage imbalance they initially induce, thereby raising the odds 
for their rapid nonfunctionalization\cite{fraser,papp,maere} (unless proved 
beneficial under concomitant environmental changes\cite{kondrashov}).  
By contrast, rapid nonfunctionalization of duplicates following a whole genome
duplication should be opposed by dosage effect. This is because  whole genome
duplications initially preserve correct relative dosage between expressed
genes, while subsequent random nonfunctionalizations disrupt this initial 
dosage balance. Preventing rapid asymmetric divergence between duplicates from
recent whole genome duplications appears, in the end, to increase their chance
of neo- or subfunctionalization by favoring longer (symmetric) genetic drift 
rather than early  (asymmetric) functional loss.

\vspace{0.1cm}

\noindent
{\bf \em Conclusion}

\vspace{-0.1cm}

Large scale topological features of PPI networks emerge ``spontaneously'' 
in the course of evolution under simple  duplication/deletion
events\cite{ispolatov1}, {\it regardless} of the specific evolutionary
advantages individual proteins might have been selected for.   
Yet, the intrinsic exponential dynamics of PPI network evolution 
by whole genome duplications (or  independent local 
duplications selected in parallel\cite{evlampiev_qbio}) {\it requires} 
an asymmetric divergence of protein duplicates. Such asymmetric divergence
arises ``naturally'' at the level of protein-binding sites or domains 
(through ``spontaneous symmetry breaking'') and is robust to extensive domain
shuffling of multi-domain proteins. 

\vspace{0.2cm}

%\footnotesize 
\noindent
{\bf Acknowledgements.} We thank
U. Alon,  M. Consentino-Lagomar\-sino, T. Fink, R. Monasson, M. Vergassola and
C. Wiggins for discussion. This work was supported by CNRS, Institut Curie and
HFSP. 

\vspace{0.1cm}

%\footnotesize 

\noindent
{\bf Correspondence:} {\tt herve.isambert@curie.fr}

\begin{widetext}

%\vspace{-0.2cm}
%\vspace{0.2cm}
\vspace{.3cm}

\noindent
%\centerline{\bf Supporting Information}
\centerline{\bf SUPPORTING INFORMATION}

%\vspace{0.2cm}
\vspace{.3cm}

\noindent
{\bf I.~~ Supplementary Figure.}

\vspace{.2cm}

%%%%%%%%%%%%%%%%%%%%%%%%%%%%%%%%%%%%%%%%%%%%%%%%%%%%%%%%%%%%%%%%%%%%%%%%%
%\begin{center}
{\centering \makebox[490pt]{ \epsfxsize=490pt
%{\centering \makebox[500pt]{ \epsfxsize=500pt
%\epsfbox{duplication_model1_z2f.eps}}}
\epsfbox{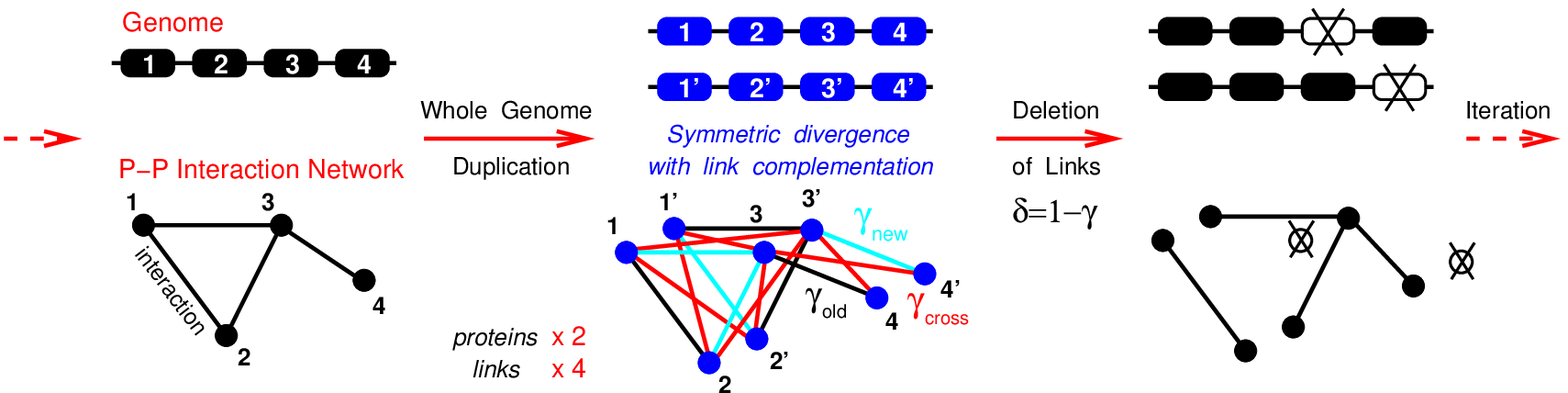}}}
\vspace{.1cm}

\noindent
{\footnotesize 
FIG.~S1:
{Alternative Model of PPI network evolution 
through whole genome duplication with
{\em symmetric
divergence} of duplicated proteins and {\em random link
``complementation''}\cite{vazquez,middendorf}. 
}}
%%%%%%%%%%%%%%%%%%%%%%%%%%%%%%%%%%%%%%%%%%%%%%%%%%%%%%%%%%%%%%%%%%%%%%%%%

%\vspace{.8cm}
\vspace{.3cm}

\noindent
{\bf II.~~ Proof of Recurrence Relations for Generating Functions (Eq.2 and Eq.15).}

\vspace{.2cm}

\noindent
After each whole genome duplication, each node has at most doubled its number
of neighbors counted through powers of $x$ in the generating function. 
Hence, a given PPI network realization  with $N_k$ nodes of connectivity $k$
($k\ge0$) will contribute to the next duplicated ensemble of PPI networks as,
\begin{equation}
%\label{}
N_kx^k \rightarrow N_kx^{2k} 
\end{equation}
After link deletion with probability $\delta$ or 
$\delta_i= \delta_o, \delta_n$, it contributes to the $x^m$ terms of
the generating function (with \mbox{$m=0,\ldots, 2k$})  as,
\begin{equation}
%\label{}
N_k x^{2k} \rightarrow N_k 
\left( \sum_{\ell=0}^k \Big( {\begin{array}{*{20}c} k \\ \ell \\
  \end{array}}\Big) (\gamma x)^\ell \delta^{k-\ell} \right)
\left( \sum_{\ell=0}^k \Big( {\begin{array}{*{20}c} k \\ \ell \\
  \end{array}}\Big) (\gamma_i x)^\ell \delta_i^{k-\ell} \right)=
N_k \Big( (\gamma x + \delta)(\gamma_i x + \delta_i)\Big)^k
\end{equation}
for the {\em asymmetric divergence} model (Fig.~1, Eq.~2) and as,
\begin{eqnarray}
%\label{}
N_k x^{2k} &\rightarrow& N_k 
\left( \sum_{\ell=0}^k \Big( {\begin{array}{*{20}c} k \\ \ell \\
  \end{array}}\Big) (\gamma x)^\ell \delta^{k-\ell} \right)
\left[ \sum_{j=0}^k \Big( {\begin{array}{*{20}c} k \\ j \\
  \end{array}}\Big) 
\left( \sum_{\ell_o=0}^j {1 \over 2^{j}}\Big( {\begin{array}{*{20}c} j \\ \ell_o \\  
  \end{array}}\Big) (\gamma_o x)^{\ell_o} \delta_o^{j-\ell_o} \right)
\left( \sum_{\ell_n=0}^{k-j} {1 \over 2^{k-j}} \Big( {\begin{array}{*{20}c}
      k-j \\ \ell_n  \\ 
  \end{array}}\Big) (\gamma_n x)^{\ell_n} \delta_n^{k-j-\ell_n}
\right)\right]\nonumber \\
&\rightarrow&  N_k \Big( (\gamma x + \delta)({\gamma_e}x + {\delta_e})\Big)^k
\end{eqnarray}
with $\gamma_e=(\gamma_o +\gamma_n)/ 2$ and $\delta_e=(\delta_o +\delta_n)/ 2$
for the {\em symmetric divergence} model with {\em link ``complementation''}\cite{vazquez,middendorf} (Fig.~S1, Eq.~15).

\vspace{.5cm}

\end{widetext}

%\newpage

\noindent
{\bf III.~~ Gene functionalization patterns in different models of 
PPI network evolution through whole genome duplication.}

\vspace{.2cm}

The initial model depicted on Fig.~1 with 
{\em asymmetric divergence} of duplicated proteins leads typically to
  ``neofunctionalization'' of ``new'' duplicates, 
while ``old'' duplicates retain most initial interactions  (if not all for
$\gamma_o=1$). 

By contrast, the alternative model depicted on Fig.~S1 with 
{\em symmetric divergence} of
 duplicated proteins and {\em random link
   ``complementation''}\cite{vazquez,middendorf} 
leads typically to random  ``subfunctionalization'' 
between protein duplicates  {\em at the level of individual interactions}.
However, this eventually leads to
exponential degree distributions with {\it no} topology conservation of the PPI
network (see main text), whereas
scale-free degree distributions with at least local topology conservation 
of the PPI network indeed emerge under the initial asymmetric model, 
Fig.~1.  

Yet, as discussed in the main text, the necessary {\em asymmetric divergence}
of protein duplicates occurs ``spontaneously'' at the level of protein-binding
sites rather than of the entire (multi-domain) proteins, as assumed in Fig.~1.
This motivates the redefinition of the initial model in terms of
protein-binding domains (Fig.~3A) to capture the
{\em asymmetric divergence} of protein duplicates {\em at the level of
  protein-binding sites} and allow, at the same time, for extensive domain
shuffling events of multidomain proteins (see main text).

This more elaborate model of PPI network evolution by whole genome
duplication and domain shuffling encompasses both
``neofunctionalization''  and   ``subfunctionalization'' of gene duplicates
{\em at the level of protein domains}, in agreement with the
suggestion that gene/protein evolution
should be analyzed in terms of domains rather than entire
proteins\cite{doolittle1,riley,koonin,apic,orengo}.
In addition, this combined model of PPI network evolution also provides a 
theoretical framework to describe the evolution of the ``combinatorial logic'' 
behind indirect interactions within multi-protein complexes (see Fig.~3A and 
main text).

\vspace{.3cm}

\noindent
{\bf IV.~~ Statistical weighting of indirect interactions from protein complexes.}

\vspace{.2cm}

We use a statistical implemention of the ``combinatorial logic'' underlying
{\it indirect} protein interactions. Indirect interactions between protein
pairs are weighted by the product of binding site ``availabilities'' along the
shortest weighted path of intermediate  direct interactions connecting
them. The ``availability'' $a_i$ of a binding site $i$ is defined as the
relative expression level ($e_i$) with respect to its first neighbor binding
partners $j$ of connectivity $d_j$, 
\begin{equation}
a_i =  {e_i \over e_i + \sum_{j\in \langle i \rangle} e_j/d_j} <1
\end{equation}
Where expression level $e_j$ can be distributed with specific statistics,
such as randomly, uniformly or according to characteristic power laws,  as
reported experimentally\cite{fraser,krylov,ueda,lemos1,lemos2}. 
Yet, in practice, we found that the predicted large scale topological features 
of PPI networks depend only weakly on the specific distribution of
expression levels (for reasonable distribution range). 

The {\it statistical probability} of an (intermediate) direct interaction 
between domains $i$ and $j$ is then proportional to $a_ia_j$, which we use 
in a Dijkstra-like algorithm\cite{dijkstra} for additive distance minimization 
assigning $d^\circ_{ij}=-\ln(a_ia_j)>0$ weights between interacting 
domains $i$ and $j$.
Because of the presence of both covalent peptide bonds and
direct, noncovalent interactions  between protein domains (Fig.~3A),
indirect protein-protein interactions correspond to {\em alternating paths} of
noncovalent and covalent interactions {\em with no successive noncovalent 
interactions} which are forbidden by the shared binding site constraint 
({\it i.e.} a binding site can only interact with one binding partner at a
time). We describe below an algorithm which performs a simultaneous 
minimization 
for paths starting with a covalent bond   ($c_{ij}$) and paths starting with
a direct, noncovalent interaction   ($d_{ij}$).  (An additional variable for 
second node $v_{ij}$ on the path is also needed to avoid non-physical
``covalent loops''.)

\noindent
The initialization of distances between protein domains is:
\begin{eqnarray}
\label{gij1}
& &c^{\circ}_{ij}={\rm Max,}\;\;\; v^{\circ}_{ij}=j  {\rm
  ~~~~~~~~for~ all~}(i,j) {\rm ~pairs,~ and}\nonumber\\
& &\delta_{ij}=d^\circ_{ij}\!=\!-\ln(a_ia_j) {\rm ~~~~~~for~ direct,~ noncovalent~ interactions,}\nonumber\\
& &\delta_{ij}=0,\;\; d^{\circ}_{ij}={\rm Max} {\rm
  ~~~~~~~~for~ covalent~ bonds,} \nonumber\\
& &\delta_{ij}=d^\circ_{ij}={\rm Max} {\rm ~~~~~~~~~~~~~~otherwise.}\nonumber 
\end{eqnarray}

\noindent
We then iterate until convergence (after $N^2\times$ (longest path)
operations): 
\begin{eqnarray}
\label{gij2}
& &d^\prime_{ij}=\min\bigl(d_{ij},{\min_{k\in \langle i \rangle_d}(\delta_{ik}+c_{kj})}\bigr)\nonumber\\
& &c^\prime_{ij}=\min\bigl(c_{ij},{\min_{k\in \langle i \rangle_c, v_{kj}\neq i}(\delta_{ik}+\min(d_{kj},c_{kj}))}\bigr)\nonumber\\
& &v^\prime_{ij}=\{k\in \langle i \rangle_c \;\vert\; v_{kj}\neq i,\;
{\min(\delta_{ik}+\min(d_{kj},c_{kj}))}\} \;\;\;\;\;\;\;\;\;\;\;\;\;\;\nonumber
\end{eqnarray}

\noindent
and remove eventually the minimum paths starting with a covalent bond
(to avoid double counting of indirect interactions for multidomain proteins
below): 
\begin{equation}
%\label{gij2}
d_{ij}={\rm Max} {\rm ~~~~~~~~~~~~~~~~~~~~~~~~if~~} d_{ij}\ge\min(c_{ij},c_{ji}){\rm ~~~~~~~~~~~~~~} \nonumber 
\end{equation}

\noindent
Hence, the probabilities to observe a {\it single indirect} interactions within protein
complexes is given by: 
\noindent
\begin{eqnarray}
\label{wij}
& &w_{ij}=0 {\rm ~~~~~~~~~~~~~~~~~~~~~~~~~~~~~~if~~~} d_{ij}={\rm Max} {\rm ~~~~~~~~~~~~~~~~~~~~~~}\nonumber\\
& &w_{ij}={\beta \exp(-d_{ij})} {\rm ~~~~~~~~~~~~otherwise,~~~~~~~~~~~~~~~~~~~~~~~~~~~~~~~~~~~~~~~~~~~~~~~~~~~~~~~~~~~~~~~~~~~~~~~}\nonumber
\end{eqnarray}
with the normalization condition $\Sigma_{i<j}w_{ij}=1$, which gives $1/\beta =
\Sigma_{i<j}\exp(-d_{ij})$.\\
$w_{ij}$ is thus the normalized product of availabilities $a_k$ along the
shortest weighted path between $i$ and $j$.\\

\noindent
Finally, the individual probabilities $p_{ij}$ to observe a total of 
$M$ {\it indirect} interactions within protein complexes are given by: 
\noindent
\begin{equation}
\label{wij}
p_{ij}=1-(1-w_{ij})^n {\rm ~~~~~~~~~~~~~~~~~~~~~~~~~~~~~~~~~~~~~~~~~~~~~~~~~~~~~~~~}
\end{equation}
where $n$ is solution of $\Sigma_{i<j}p_{ij}=M$.\\

\noindent
Given the number $M$ of indirect interactions in various data
sets\cite{gavin2002,ho2002,gavin2006}, we have assessed their 
expected contribution to the large scale topology of Yeast PPI network 
from the two-parameter  $\gamma-\lambda$ model
described in the main text. 
$M\simeq 28,000$ corresponds to the sum of about $9,000$ direct physical
interactions from the BIND database\cite{bind} (Fig.~2B\&C filled symbols) 
and about  $19,000$ ``matrix'' interactions from
\cite{ho2002,gavin2002}  between 
$2,100$ proteins already involved in direct physical
interactions (out of $4,576$ proteins in the BIND database, Fig.~3C filled 
symbols). ``Matrix'' interactions 
from ref.\cite{gavin2006} (Fig.~3C open symbols) are ``reconstructed'' from 
supplementary information files of\cite{gavin2006} as follows: ``matrix''
interactions are included for (each complex core)$\times$(each associated
``module'') and  (each complex core)$\times$(each associated
``attachment'' = one protein). This reconstructed dataset should therefore be
considered as incomplete, since ``matrix'' interactions between compatible
modules and/or attachments  associated to a given core are {\it not}  taken
into account (information {not} 
given in\cite{gavin2006}). \\

\noindent
Numerical fits ($\gamma=0.1$, $\lambda=0.3$) 
are displayed on Fig.~3C (for  direct {\em and} indirect interactions)
for both connectivity distribution (green) and average connectivity of first
neighbors (blue).
They corresponds to
the {\em same} adjusted values ($\gamma=0.1$, $\lambda=0.3$) as in Fig.~3B 
(for direct interactions only).\\

\vfill
\eject

\end{document}